\documentclass[manuscript]{aastex}

\usepackage{lscape}

\slugcomment{Accepted for publication in the Astrophysical Journal}

\shorttitle{The Early ALMA view of the FU~Ori outburst system}
\shortauthors{Hales et al.}

\begin{document}


\title{The Early ALMA View of the FU~Ori Outburst System}



\author{A. S. Hales \altaffilmark{1,2}, S. A. Corder \altaffilmark{1,2}, W. R. D. Dent \altaffilmark{1,3}, S. M. Andrews \altaffilmark{4}, J. A. Eisner \altaffilmark{5} and L.~A.~Cieza \altaffilmark{6,7}}

\altaffiltext{1}{Atacama Large Millimeter/Submillimeter Array, Joint ALMA Observatory, Alonso de C\'ordova 3107, Vitacura 763-0355, Santiago - Chile}
\altaffiltext{2}{National Radio Astronomy Observatory, 520 Edgemont Road, Charlottesville, Virginia, 22903-2475, United States}
\altaffiltext{3}{European Southern Observatory, Karl-Schwarzschild-Str. 2, 85748, Garching bei M\"unchen, Germany}
\altaffiltext{4}{Harvard-Smithsonian Center for Astrophysics, Cambridge, MA 02138, United States}
\altaffiltext{5}{Steward Observatory, University of Arizona, 933 N Cherry Ave., Tucson, AZ 85721, United States}
\altaffiltext{6}{N\'ucleo de Astronom\'ia, Facultad de Ingenier\'ia, Universidad Diego Portales, Chile}
\altaffiltext{7}{Millenium Nucleus, Protoplanetary Disks in ALMA Early Science, Chile}
\email{ahales@alma.cl}



\begin{abstract}


We have obtained ALMA Band 7 observations of the FU~Ori outburst
system at 0.6$\arcsec\times$0.5$\arcsec$ resolution to measure the
link between the inner disk instability and the outer disk through
sub-mm continuum and molecular line observations. Our observations
detect continuum emission which can be well modeled by two unresolved
sources located at the position of each binary component.  The
interferometric observations recover the entire flux reported in
previous single-dish studies, ruling out the presence of a large
envelope. Assuming that the dust is optically thin, we derive disk
dust masses of $2\times 10^{-4}$M$_{\odot}$ and $8\times
10^{-5}$~M$_{\odot}$, for the north and south components
respectively. We place limits on the disks' radii of $r<$45~AU. We
report the detection of molecular emission from $^{12}$CO(3-2),
HCO$^{+}$(4-3) and from HCN(4-3). The $^{12}$CO appears widespread
across the two binary components, and is slightly more extended than
the continuum emission. The denser gas tracer HCO$^{+}$ peaks close to
the position of the southern binary component, while HCN appears
peaked at the position of the northern component. This suggests that
the southern binary component is embedded in denser molecular
material, consistent with previous studies that indicate a heavily
reddened object. At this angular resolution any interaction between
the two unresolved disk components cannot be disentangled. Higher
resolution images are vital to understanding the process of star
formation via rapid accretion FU~Ori-type episodes.

\end{abstract}





\section{Introduction}

It is well understood that stars acquire most of their material
through an accretion disk that is fed, at least at early times, by a
surrounding envelope.  Once the bulk of the envelope is dissipated,
the accretion rates tend to approach $\sim 10^{-7}$ or
$10^{-8}$~M$_{\odot}$~yr$^{-1}$.  A small percentage of young stars --
FU~Orionis or FU~Ori objects -- exhibit rapid optical brightening of
several magnitudes over timescales as short as one year
\citep{1966VA......8..109H}.  During the outburst phase, the accretion
rates increase by factors of $10^{3}$ or more relative to the
quiescent periods, reaching rates as high as
$10^{-4}$~M$_{\odot}$~yr$^{-1}$.  This high state can last for
$\sim$100 years and deposit a significant amount of disk material
($\sim$0.01\,M$_\odot$) onto the young star in a short period
\citep[][ and references therein]{1996ARA&A..34..207H}.

%


The cause of these outbursts is not well understood. Different
mechanisms have been proposed as possible origin for the outburst \citep{2014prpl.conf..387A}: 1)
thermal instability \citep{1994ApJ...427..987B}, 2) coupling of
gravitational and magnetorotational instabilities
\citep{2001MNRAS.324..705A,2009ApJ...694L..64Z}, 3) disk fragmentation
\citep{2006ApJ...650..956V,2012ApJ...746..110Z}, and 4) tidal
interactions between the disk and either planets
\citep{2004MNRAS.353..841L} or stellar companions
\citep{1992ApJ...401L..31B}. 


Connecting infrared studies which probe inside radii of 1-10\,AU to
the main mass reservoirs in the outer disk and/or envelope is
particularly important if a full understanding of the phenomenon is to
be achieved.  For instance, it remains unclear whether the reservoir
of material feeding the enhanced accretion rate is distributed in a
circumstellar disk, in a larger envelope, or both. The only way to
accurately recover the surface density distribution of the gas and
dust around FU~Ori objects is via resolved imaging of the disk at,
preferably multiple, (sub)millimeter wavelengths and/or transitions
where the emission is most likely optically thin.

If the disk is small, and if there is no reservoir of material from
which to replenish it, if it continues to accrete at
$10^{-4}$M$_{\odot}\,$yr$^{-1}$, the disk may be quickly dying.  If
the disk is indeed more massive and/or larger than previous
submillimeter observations indicate, it is then a counterpart to the
more massive disks seen in nearer star forming regions and we are
simply seeing one outburst of potentially many.  Clearly, resolving
the source and determining the mass is key.

\section*{FU Ori}\label{source}

Despite being the prototype of the FU~Ori class and one of the nearest
such objects, at a distance of 450\,pc \citep{1977MNRAS.181..657M},
little is known about the global properties of the accretion disk
surrounding FU Ori itself.  The estimated age of the system is
$\sim$2~Myr \citep{2012AJ....143...55B}. Early interferometric studies
at 1.3~mm by \citet{1995PhDT........11M} found an unresolved (at
$\sim$3.5\arcsec resolution, or $\sim$1500~AU ) disk with
$^{13}$CO(2-1) emission spanning 8.1-13.3~km~s$^{-1}$ with a total
integrated intensity of 1.4\,Jy.  Studies with the JCMT by
\citet{2001ApJS..134..115S} show an unresolved continuum source
($<$5\arcsec in diameter) and a flux of 66 and 400 \,mJy at 850 and
450\,$\mu$m, respectively. These results suggest there is no envelope
bigger than 1500~AU around the source.

Significant progress has been made towards understanding the inner
disk of FU Ori objects in general and FU Ori specifically \citep[see
  papers by e.g. ][]{2009ApJ...694L..64Z, 2011ApJ...738....9E}.  The
region of rapid accretion for FU Ori itself has been shown to be
occurring over a radius of $\sim$1\,AU, or 0.005\arcsec, in diameter.
Given the size of the region involved, thermal instability alone
cannot explain the outburst and appeals to other outbursts mechanisms 
must be made. 

FU~Ori was discovered to be a binary system by
\citep{2004ApJ...601L..83W}, with the newly discovered binary
component (FU~Ori~South, or FU~Ori~S hereafter) located 0.5\arcsec to
the South of the brighter principal component, FU~Ori. Recent studies
indicate FU~Ori~S is actually the more massive star in the system
\citep[$\sim$1.2\,M$_\odot$;][]{2012AJ....143...55B}, heavily embedded
in obscuring material \citep{2012ApJ...757...57P}. A number of FU~Ori
type objects have now been identified as having binary components
\citep[e.g., L1551~IRS5, RNO~1B/C, AR~6A/B][ and references
  therein]{2012ApJ...757...57P}, triggering discussion on whether the
rapidly accreting FU~Ori phenomenon could be caused by close-companion
interaction. However, at least in the case of FU~Ori itself, this
hypothesis remains inconclusive \citep[e.g. see the discussion in
][]{2012AJ....143...55B}.


\citet{2001ApJS..134..115S} suggest an upper limit for the disk mass
at 0.02\,M$_\odot$.  There are a variety of ways to constraint the
dust, gas and total mass of the disk, all of which involve obtaining
resolved imaging to determine the optically thick contribution and
model fits \citep{{2007ApJ...659..705A},2008ApJ...683..304E}.  Below
we present new observations of FU Ori with sufficient resolution to
constrain the masses of disks around FU Ori~N and S.


\section{Observations and data reduction}\label{obs}

FU Ori Band 7 observations were acquired on December 2$^{nd}$ 2012
using 24 antennas. The dataset provides baselines from 15 to 381
meters. During the observations the precipitable water vapor in the
atmosphere was stable between 1.3 and 1.7~mm with clear sky
conditions, resulting in median system temperature of 190~K. The ALMA
correlator was configured in the Frequency Division Mode (FDM) to
provide 468.75~MHz bandwidth in each of four different spectral
windows at 122~kHz ($\sim$0.1~km~s$^{-1}$) resolution per
channel. Three spectral windows were positioned to target the
$^{12}$CO~$J=3-2$, HCN~$J=4-3$ and HCO$^+$~$J=4-3$ transitions at
345.79599, 354.50547 and 356.73424~GHz, respectively. The fourth
spectral window was positioned in a region devoid of line emission for
detecting the continuum dust emission (centered at
345.78359~GHz). Callisto was observed as flux calibrator, while the
quasars J0522-364 and J0532+075 were observed for bandpass and phase
calibration respectively (J0532+075 is located 3.5~degrees away from
the science target). Observations of the phase calibrator were
alternated with the science target every 5 minutes to calibrate the
time-dependent variations of the complex gains. The total time spent
on-source was 16.9 minutes. A secondary phase calibrator,
J053056+13322, was also observed regularly (located 5.7~degrees away).

All the data were calibrated using the {\it{Common Astronomy Software
    Applications }} package
\citep[CASA{\footnote{\url{http://casa.nrao.edu/}}};][]{2007ASPC..376..127M}
in a standard fashion, which included offline Water Vapor Radiometer
(WVR) calibration, system temperature correction, as well as bandpass,
phase and amplitude calibrations.  After flagging for problematic
antennas, a total of 21 antennas were used for imaging. The fluxes
derived for J0522-364, J0532+075 and J053056+133220 were 7.5, 1.1 and
0.5~Jy respectively. During Cycle 0, the absolute flux calibration of
the data is estimated to be accurate to within 10 to 15$\%$.

Imaging of the continuum and molecular emission lines was performed
using the CLEAN task in CASA. Using natural weighting resulted in a
synthesized beam size of 0.63$\arcsec\times$0.51$\arcsec$ at
PA$=3.0$~degrees. Continuum subtraction in the visibility domain was
performed prior to imaging of each molecular line. After CLEANing the
images, an RMS noise level of 25.0~mJy\,beam$^{-1}$ per 0.22~km~s$^{-1}$ channel
was reached (in the $^{12}$CO~$J=3-2$ line).  CLEANing of the dust
continuum was performed after combining the line-free channels from
all four spectral windows (adding to a total bandwidth of 1.759~GHz
centered at 351.33~GHz), for which an RMS of 0.64~mJy\,beam$^{-1}$ was
obtained. After a single iteration of phase-only self-calibration the
RMS of the continuum image was reduced to 0.31~mJy\,beam$^{-1}$.

\section{Results}\label{results}

\subsection{Dust Continuum}\label{cont-results}

\begin{center}
\begin{figure}
\includegraphics[angle=0,scale=0.33]{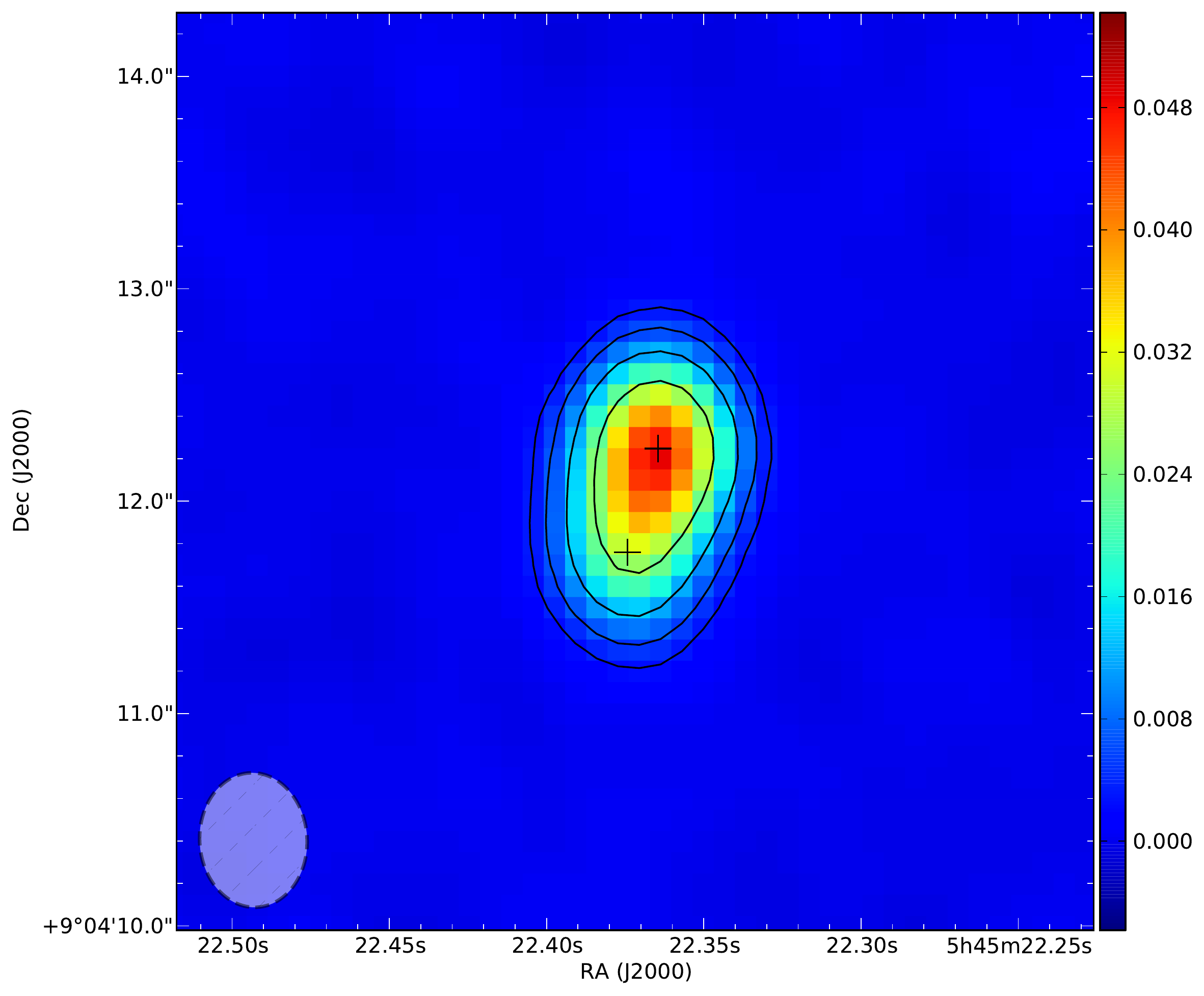}
\includegraphics[angle=0,scale=0.33]{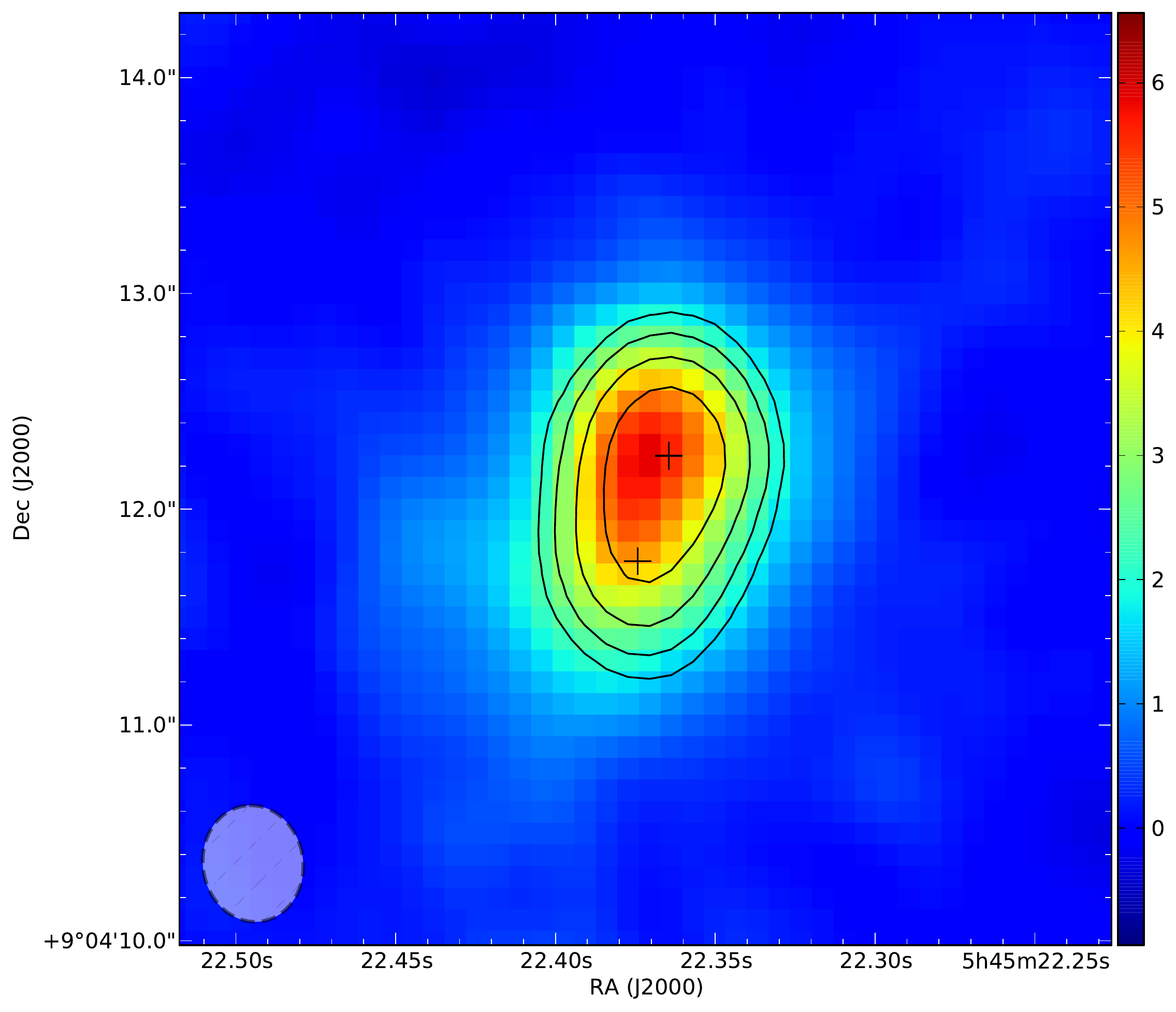}

\caption{{\it{Left-panel:}} ALMA Band 7 continuum image of the FU~Ori
  system with contours overlayed (after self-calibration). Contour
  levels are 10, 20, 40 and 80$\times$0.31~mJy\,beam$^{-1}$ (the rms
  noise). The stellar position from 2MASS is shown with a black star
  symbol. {\it{Right-panel:}} $^{12}$CO(3-2) integrated intensity
  (moment 0) image with Band 7 continuum contours overlayed (Contour
  levels are identical in both figures). The peak $^{12}$CO(3-2)
  flux is 5.8~Jy~beam$^{-1}$~km~s$^{-1}$.  }\label{fig-1}
\end{figure}
\end{center}

Figure~\ref{fig-1} shows the ALMA Band 7 continuum image of FU~Ori
after self-calibration. The continuum emission at 0.854~mm
(351.33~GHz) is detected at a signal-to-noise ratio (SNR) of
$\sim$150.  The emission is resolved, with the peak roughly coinciding
with the position of the northern binary component (FU Ori). We used
the CASA task IMFIT to fit the observed emission using a single
component 2-D Gaussian, but inspection of the residual image shows
that a single-component fit fails to match data. Using two unresolved
sources provides a much better fit to the observed emission (with
residuals down to the image noise levels). The total fluxes of each
component are 50.1$\pm 0.3$~mJy and 21.2$\pm 0.4$~mJy respectively.

\noindent The parameters of the Gaussian components determined using
IMFIT are detailed in Table~\ref{table-1}. The positions of each
Gaussian component are shown in Figure~\ref{fig-1}. The position of
the brighter component coincides within 0.1'' of the optical position
of FU~Ori (as listed in 2MASS), whilst the fainter disk is located
0.5'' to the South-East (0.48'' to the South and 0.14 to the
East). This is similar to the separation between the FU~Ori and
FU~Ori~S, as reported in previous near-IR observations \citep[e.g. ][
  and references
  therein]{2004ApJ...601L..83W,2009ApJ...700..491M,2012AJ....143...55B,2012ApJ...757...57P}.


\subsection{Line emission}\label{gas-results}

\begin{center}
\begin{figure}
\includegraphics[angle=0,scale=0.33]{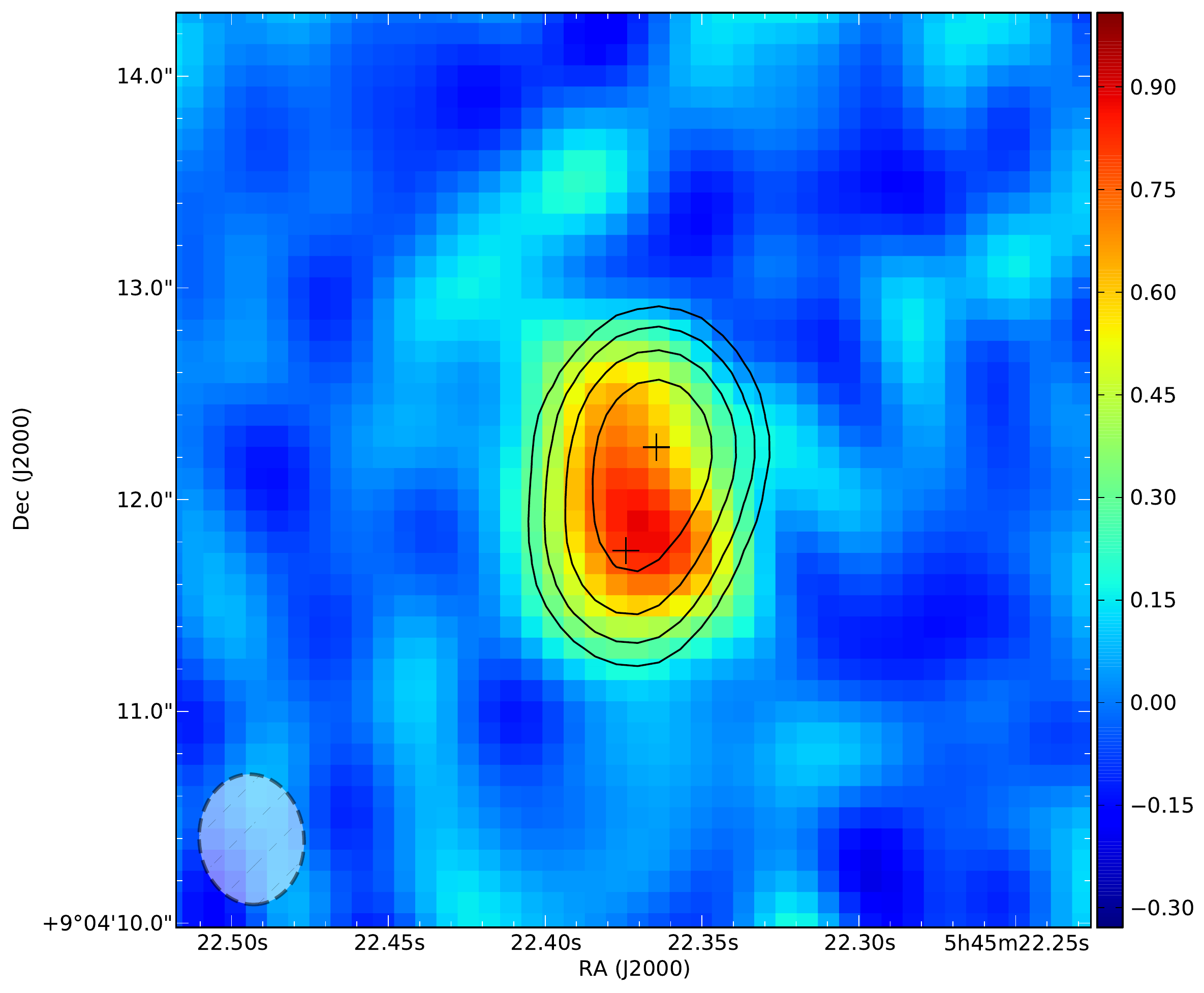}
\includegraphics[angle=0,scale=0.33]{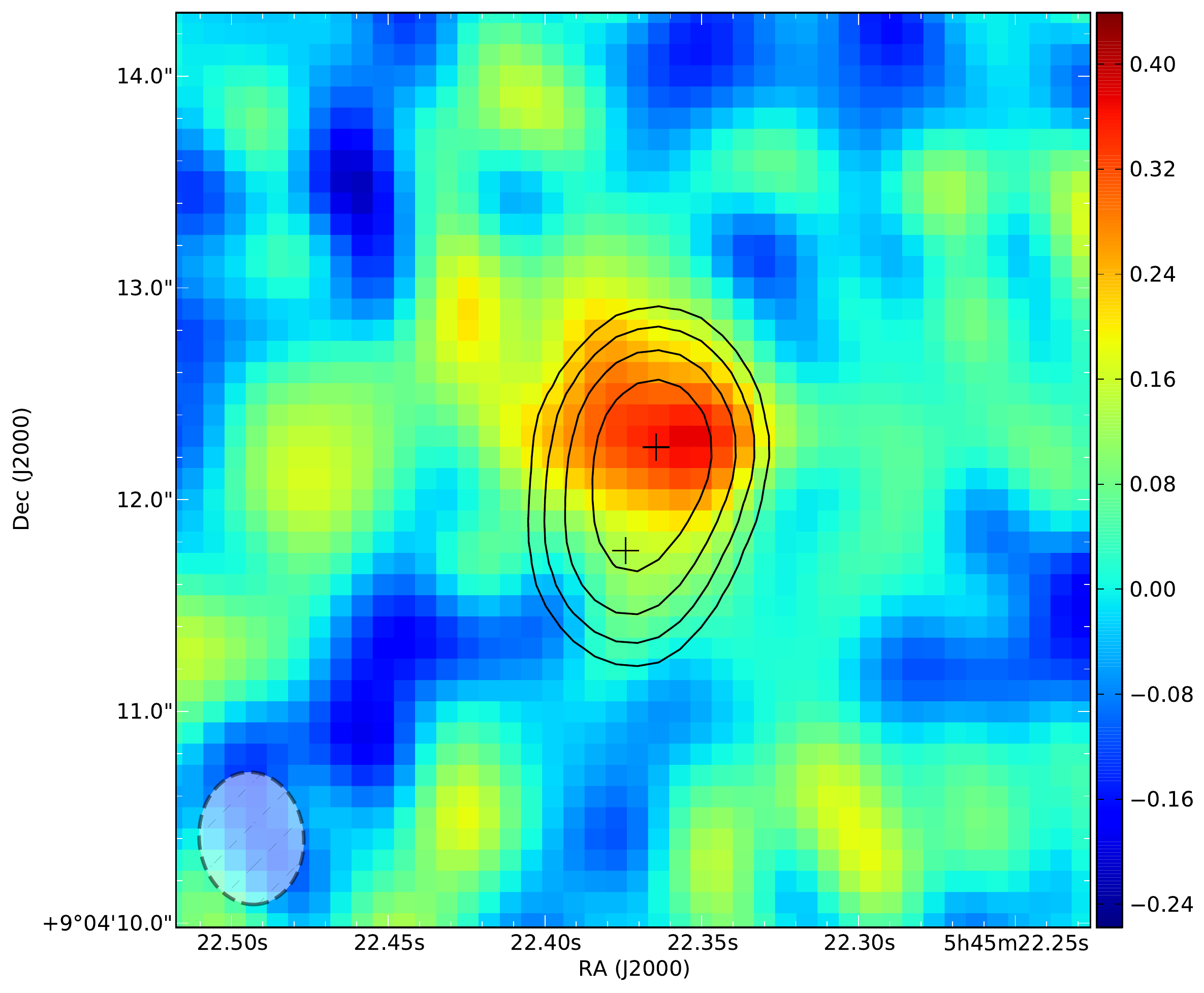}

\caption{{\it{Left-panel:}} HCO$^{+}$(4-3) integrated intensity
  (moment 0) image with Band 7 continuum contours overlayed.  The peak
  flux is 0.88~Jy~beam$^{-1}$~km~s$^{-1}$. {\it{Right-panel:}}
  HCN(4-3) moment 0 image with Band 7 continuum contours
  overlayed. Contour levels are identical to Figure~\ref{fig-1}. The
  peak flux is 0.37~Jy~beam$^{-1}$~km~s$^{-1}$.  }\label{fig-2}
\end{figure}
\end{center}

\begin{center}
\begin{figure}
\includegraphics[angle=0,scale=0.7]{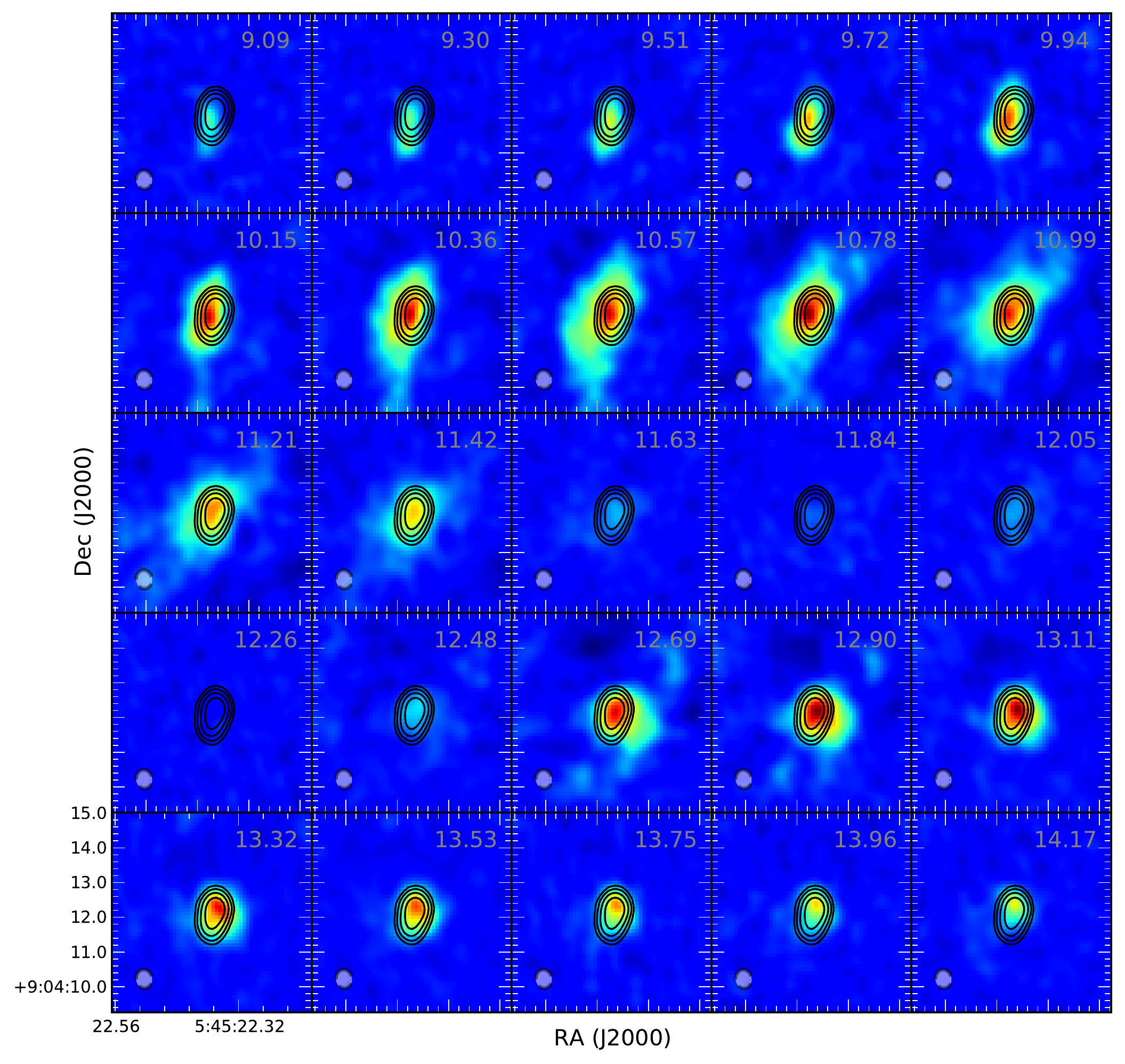}
\caption{ CO channel maps towards FU~Ori. The velocity of the channels
  is shown in the Local Standard of Rest (LSR) frame, centered at the
  rest frequency of $^{12}$CO~$J=3-2$. The data has been binned to a
  velocity resolution of 0.21~km~s$^{-1}$.  The band 7 continuum
  contours are overlayed (with contour levels are identical to
  Figure~\ref{fig-1}). Cloud contamination is present in the channels
  between 11.63 and 12.69~km~s$^{-1}$. All maps share the same linear
  color scale, ranging from -0.2 to 1.5~Jy\,beam$^{-1}$.}
\label{fig-4}
\end{figure}
\end{center}

We detect emission from $^{12}$CO~$J=3-2$, HCO$^+$~$J=4-3$ and
HCN~$J=4-3$ molecules at SNR levels (peak/rms) of 47, 12 and 5
respectively.  Figure~\ref{fig-1} and Figure~\ref{fig-2} show the
integrated emission maps for each molecule. The integrated line
emissions are 31.61$\pm 0.12$, 2.14$\pm 0.07$ and 0.37$\pm
0.07$~Jy~km~s$^{-1}$ respectively. Figure~\ref{fig-4} shows the
$^{12}$CO channel maps towards FU~Ori. The integrated intensity maps
of the three molecules are strikingly different. $^{12}$CO is more
extended compared to the continuum (extending up 1'' from FU~Ori, and
covering the two binary components), and peaks east-ward from FU~Ori.
On the other hand, HCO$^{+}$ peaks close to the position of the
southern binary component (FU~Ori~S), whilst HCN~$J=4-3$ appears to
peak closer to FU~Ori.

\begin{center}
\begin{figure}
\includegraphics[angle=0,scale=0.33]{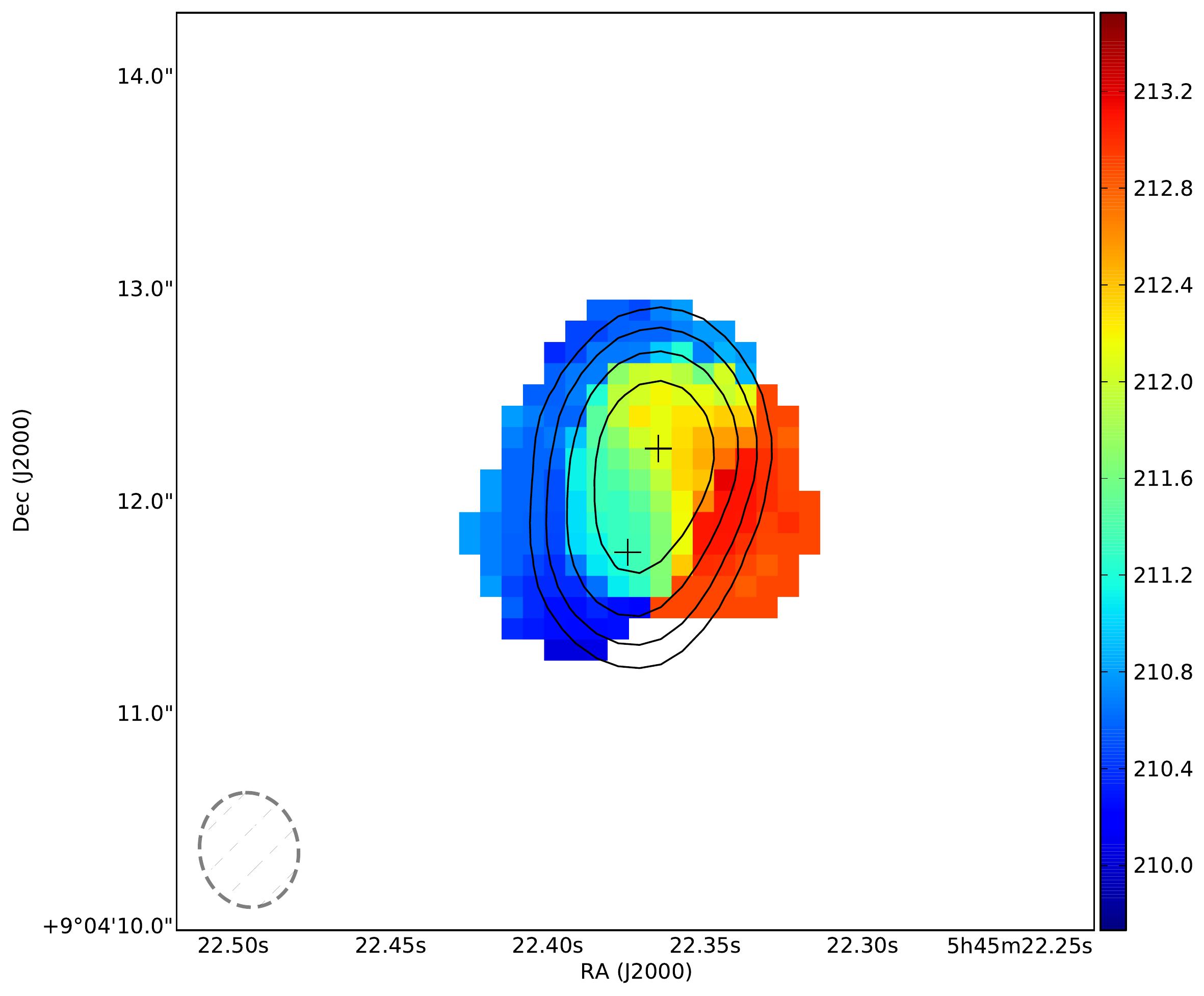}
\includegraphics[angle=0,scale=0.46]{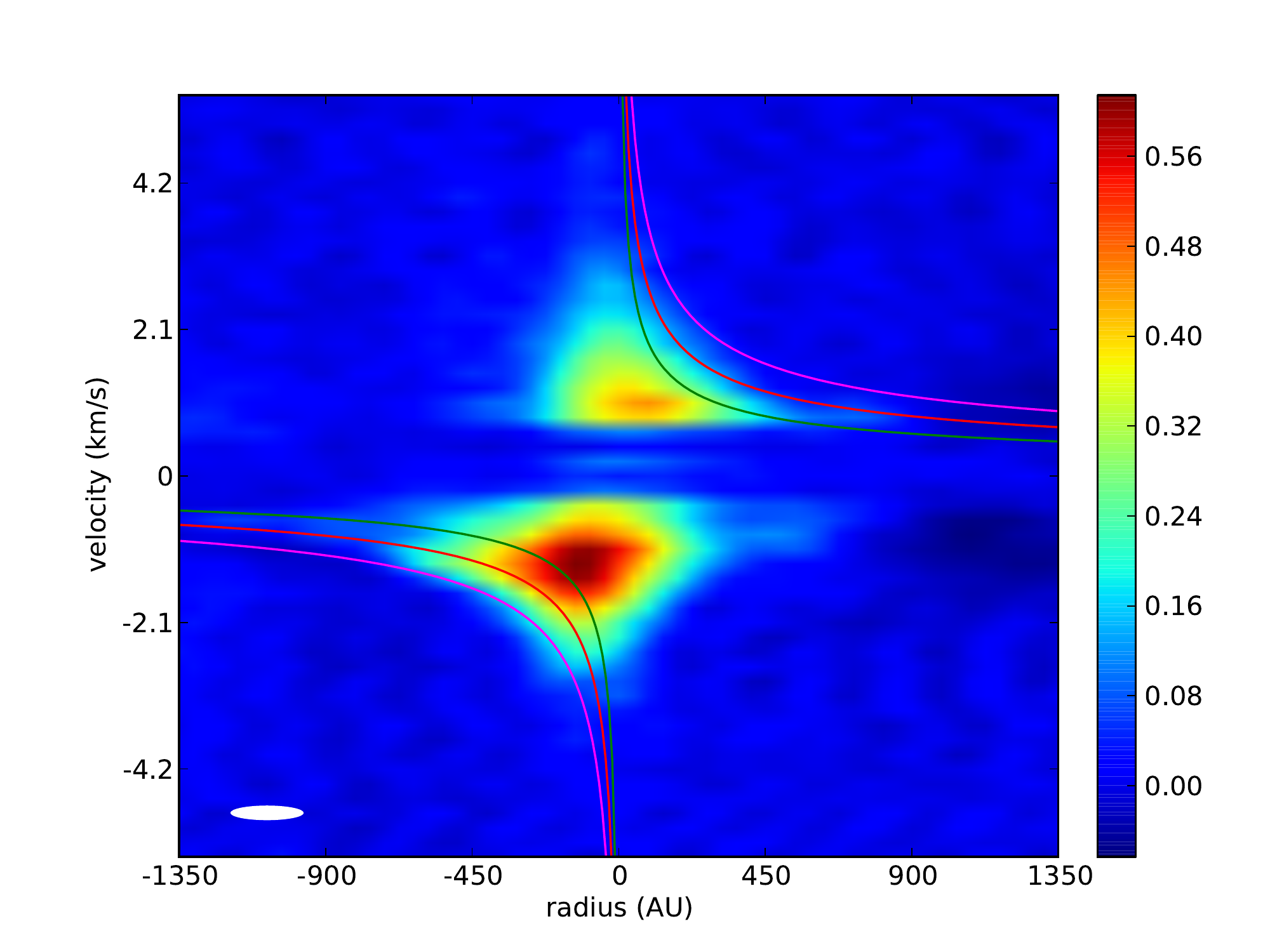}
\caption{
{\it{Left-panel:}}
CO moment 1 (intensity-weighted velocity field) of FU~Ori. The
position of the two fitted continuum point sources and the band 7
continuum contours are overlayed (with contour levels are identical to
Figure~\ref{fig-1}). 
{\it{Right-panel:}}
FU~Ori position-velocity diagram for the observed $^{12}$CO line
emission, centered at the velocity of 11.8~km~s$^{-1}$. The spectra
$\pm$1'' on either side of the apparent major axis have been averaged
to form this PV diagrams. Spectral/spatial resolution is shown in the
lower left. For comparison, the solid lines show the maximum Keplerian
velocity for gas at the tangential point (at three different
inclination angles), as a function of distance from the star.  The
mass of the central object has been assumed to be 1.5 times the mass
of the Sun (i.e. the mass of the combined binary system). Green, red
and magenta correspond to inclination angles of 30, 45 and 70~degrees
respectively.  }

\label{fig-5}
\end{figure}
\end{center}

Figure~\ref{fig-4} shows the $^{12}$CO channel maps towards
FU~Ori. There is significant cloud contamination in the central
channels, making the interpretation of the $^{12}$CO spectrum
difficult. The velocity of the channels with no $^{12}$CO emission
(11.8~km~s$^{-1}$) is very similar to the velocity channels where
strong $^{12}$CO(1-0) emission was previously detected with the NRAO
11m telescope \citep{1982ApJ...259L..35K}. This suggests that
unresolved emission and/or absorption from the larger Orion molecular
clouds contaminates the emission in the central channels of the ALMA
spectrum. Nevertheless, a velocity gradient can be perceived in the
channel maps and in the first moment map of the $^{12}$CO
(Figure~\ref{fig-5}).  Figure~\ref{fig-5} shows the position-velocity
diagram taken in the East-West direction (where the velocity gradient
appears to be larger), which suggests there is gas in keplerian
rotation around the system.


\section{Discussion}\label{disc}

We detect continuum emission towards both binary components,
suggesting they both harbour circumstellar disks. The disks are
unresolved at the $\sim$0.6$\arcsec\times$0.5$\arcsec$ resolution of
our observations (or 225~AU at FU~Ori's 450~pc distance), placing
upper limits on the disk radial sizes. To be consistent with the
Gaussian fit, the disks must have radial sizes smaller than 0.1''
(45~AU). At the projected separation of the binary system (225~AU),
the disks radii are predicted to be truncated at 0.2 to 0.5 times the
binary separation \citep{1996ApJ...467L..77A}. The deconvolved disk
sizes of 45~AU coincide with 0.2 times the binary separation.

Our ALMA data rule out the presence of a large envelope around the two
stars, since our observations would have detected any extended
emission smaller than $7$'' (3000~AU) in diameter. Our observations
also place limits on the presence of a possible circumbinary ring
orbiting the system. Assuming a 0.5'' wide ring, starting at 1.5''
\citep[given the minimal truncation radius for the projected binary
  orbit; e.g.][]{1996ApJ...467L..77A}, the ring would need to have a
dust mass smaller than $1.9\times 10^{-4}$M$_{\odot}$ to remain
undetected by our observations (at 3-$\sigma$, assuming a dust
temperature of 20~K). This is similar to recent ALMA results from
\citet{2014ApJ...784...62A} in which no circumbinary emission was
detected in any of the 17 binary stars of their sample.

Assuming a dust temperature of 50~K for the individual disks around
FU~Ori and FU~Ori~S, a gas-to-dust ratio of 100, and a 345~GHz dust
opacity of 1.5~cm$^2\,$g$^{-1}$, we can compute a coarse estimate of
the disks' total masses \citep[e.g. ][]{{1990AJ.....99..924B}}. Under
these assumptions, the ALMA 854~$\mu$m fluxes correspond to disk dust
masses of $2\times 10^{-4}$~M$_{\odot}$ and $8\times
10^{-5}$~M$_{\odot}$, for the north and south components respectively.
The total mass of the combined disks is comparable to the mass
estimated from unresolved observations
\citep{2001ApJS..134..115S}. These mass estimates are, however,
subject to the assumption that the dust emission is optically thin.
For the disks to be optically thick, the average disk dust temperature
would have to be higher than 80~K. Assuming a $R^{-0.5}$ dependence
for the disk temperature, an average T~$>$ 80~K is reached if the
temperature at 45~AU is greater than 60~K. This is high for a typical
T~Tauri disk, but it corresponds roughly to the radiative equilibrium
temperature for a source with a luminosity larger than about
4~L$_{\odot}$.  This is perfectly feasible for FU~Ori, where the
outburst luminosity much larger than the photospheric luminosity. We
conclude that the disks could be marginally optically thick, although
higher resolution imaging would be required to accurately constrain
this.


According to \citet{2012AJ....143...55B} the masses of FU~Ori and
FU~Ori~S are 0.3~M$_{\odot}$, and 1.2~M$_{\odot}$ respectively.
Assuming an effective temperature of 4000-6500~K and an optical
extinction of A$_V$=8-12, \citet{2012ApJ...757...57P} estimate a
minimum mass for FU~Ori~S of 0.5~M$_{\odot}$. Both studies agree that
FU~Ori~S is the most massive star in the system. In this scenario
FU~Ori~S would be the primary star in the system. Unfortunately, our
line data lacks the angular resolution to resolve the velocity
gradients across each disk, which could be used to confirm these
stellar masses.


Theoretical models predict that the circumprimary disks should
be the most massive disks in the system
\citep[e.g. ][]{2000MNRAS.314...33B}. However, recent submillimeter
surveys of binary systems show no clear relationship between disk
masses and stellar mass ratios
\citep{2012ApJ...751..115H,2014ApJ...784...62A}.  As noted by these
authors, the predictions from these theoretical models only consider
the initial conditions of the binary disks and do not take into
account the dynamical evolution of the disks (e.g. viscous accretion
and other dissipative processes). Assuming that both disks have the
same average dust temperature, we find that in the FU~Ori system the
brightest and most massive disk orbits the less massive star. We note
that this scenario may be different if the southern disk is cooler
than the northern disk. For instance, if the southern disk has an
average dust temperature of 20~K, its mass would be higher than the
$2\times 10^{-4}$~M$_{\odot}$ dust mass derived for the northern
disk.


The denser gas tracer HCO$^+$~$J=4-3$ peaks in the position of the
most massive star FU~Ori~S, and is not detected in the direction of
FU~Ori.  Since HCO$^+$ is a higher density gas tracer, this suggests
that FU~Ori total disk mass is probably lower than FU~Ori~S's disk. On the
other hand, HCN is detected only toward FU~Ori. HCN forms at large
heights above the midplane, where the temperature are higher and the
densities are lower compared to the midplane
\citep[e.g.][]{2010ApJ...722.1607W}. Therefore, the HCN detection around
FU~Ori and not FU~Ori~S's could support the idea that the disk around
FU~Ori has a higher dust temperature compared to the disk around
FU~Ori~S. As noted above, an average dust temperature below 20~K for
FU~Ori~S would make it the most massive of the two disks, which would
be in agreement with the interpretation of the HCO$^+$ detection.


Based on the stellar photospheric absorption features
\citet{2012AJ....143...55B} estimate the accretion rate of FU~Ori~S in
$(2-3)\times 10^{-8}$M$_{\odot}$\,yr$^{-1}$. For FU~Ori the accretion
rate derived by \citet{2007ApJ...669..483Z} is $2\times
10^{-4}$M$_{\odot}$\,yr$^{-1}$. However this is highly model dependent
is likely to correspond to the maximum accretion rate during the peak
periods of episodic accretion. It is unlikely that FU~Ori's high
accretion rate can continue for long. If FU~Ori continues to accrete
at $10^{-4}$~M$_{\odot}$~yr$^{-1}$, its submillimeter flux would
decrease by a factor of ten within a decade. To sustain the disk dust
mass we infer from our ALMA observations, the stellar accretion rate
must be highly variable.

\section{Conclusions}\label{conc}

We obtained ALMA Band 7 observations of the FU~Ori binary system at
0.6$''\times$0.5$''$ resolution in Cycle 0. Our observations detect
continuum emission from two unresolved sources at the position of each
binary component. This indicates that both binary components have
disks. Under simple assumptions, these observations suggest that the
most massive disk orbits the less massive star (FU~Ori).

The interferometric observations recover the entire flux reported in
previous single-dish studies, ruling out the presence of a large
envelope for replenishing the disks. The dust disks were unresolved in
our observations, implying that they are significantly smaller than
non FU~Ori objects such as e.g. AB~Aur, HL~Tau and HD~163296.


We report the detection of molecular emission from $^{12}$CO(3-2),
HCO$^{+}$(4-3) and from HCN(4-3). Whilst the $^{12}$CO appears
widespread across the two binary components , HCO$^{+}$ peaks close to
the position of the southern binary component. This suggests that the
southern binary component is embedded in dense molecular material,
which is consistent with infrared studies that indicate that this is a
heavily reddened object. Is this clump the reservoir of material
feeding the outburst of the northern binary component? At this angular
resolution the interaction between the two unresolved disk cannot be
disentangled.

Observations at higher spatial resolution are mandatory to investigate
whether interactions between the disks or other mechanisms (e.g. disk
fragmentation) are causing the outburst.

\section*{Acknowledgments}

This paper makes use of the following ALMA data:
ADS/JAO.ALMA.2011.0.00538.S. ALMA is a partnership of ESO
(representing its member states), NSF (USA) and NINS (Japan), together
with NRC (Canada) and NSC and ASIAA (Taiwan), in cooperation with the
Republic of Chile. The Joint ALMA Observatory is operated by ESO,
AUI/NRAO and NAOJ. The National Radio Astronomy Observatory is a
facility of the National Science Foundation operated under cooperative
agreement by Associated Universities, Inc. This research made use of
Astropy, a community-developed core Python package for Astronomy
\citep{2013A&A...558A..33A}. JAE acknowledges support from NSF AAG
grant 1311910. L.A.C. was supported by ALMA-CONICYT grant number
31120009 and CONICYT-FONDECYT grant number 1140109. L.A.C. also
acknowledges support from the Millennium Science Initiative (Chilean
Ministry of Economy), through grant ''Nucleus RC130007''.

{}

%
%
%
%
%

\begin{table}
  \caption{Integrated Fluxes and Gaussian Fit Parameters}
  \label{table-1}
  \begin{center}
    \leavevmode
    \begin{tabular}{lcc} \hline \hline                 
Component  & Gaussian 1 &  Gaussian 2 \\
\hline \hline   
Position (Ra, Dec)     & (05:45:22.36,+09.04.12.24) & (05:45:22.37,+09.04.11.75) \\
Major axis (mas)     & 662.3 $\pm$ 3.4   & 636.3 $\pm$ 9.0\\
Minor axis (mas)     & 536.8$\pm$4.3     & 517.9 $\pm$ 8.9  \\
Position angle (deg)   &  0.8 $\pm$ 0.7  & 9.9 $\pm$ 2.3  \\
Peak Intensity (mJy\,beam$^{-1}$)   & 45.2$\pm$0.3      & 20.9 $\pm$ 0.3\\
Integrated  Flux (mJy) & 50.1$\pm$ 0.3     & 21.2 $\pm$ 0.4\\
 \hline  
    \end{tabular}
  \end{center}
\end{table}


\end{document}